\documentclass[aps,10pt,twocolumn,superscriptaddress]{revtex4-1}
\usepackage[ascii]{inputenc}
\usepackage{amsmath,amssymb,amsfonts,amsthm}
\usepackage{graphicx}
\usepackage[caption=false]{subfig}
\usepackage[pdftex,bookmarks=false,colorlinks=true,linkcolor=blue,
citecolor=blue,filecolor=black,urlcolor=blue]{hyperref}

\newcommand{\ud}{\mathrm{d}}

\newcommand{\N}{\mathbb{N}}
\newcommand{\Z}{\mathbb{Z}}

\newcommand{\R}{\mathbb{R}}

\newcommand{\cdag}{c^{\dagger}}
\newcommand{\cann}{c^{\vphantom{\dagger}}}
\newcommand{\chatdag}{\hat{c}^{\dagger}}
\newcommand{\chatann}{\hat{c}^{\vphantom{\dagger}}}

\DeclareMathOperator{\e}{e}

\graphicspath{{figures/}}

\begin{document}

\title{Stability of dynamical quantum phase transitions in quenched topological insulators:\\ From multiband to disordered systems}

\author{Christian B.~Mendl}
\email{christian.mendl@tum.de}
\affiliation{Technische Universit\"at Dresden, Institute of Scientific Computing, Zellescher Weg 12-14, 01069 Dresden, Germany}
\affiliation{Technische Universit\"at M\"unchen, Department of Informatics and Institute for Advanced Study, Boltzmannstra{\ss}e 3, 85748 Garching, Germany}

\author{Jan Carl Budich}
\email{jan.budich@tu-dresden.de}
\affiliation{Institute of Theoretical Physics, Technische Universit\"at Dresden, 01062 Dresden, Germany}

\date{December 26, 2019}

\begin{abstract}
Dynamical quantum phase transitions (DQPTs) represent a counterpart in non-equilibrium quantum time evolution of thermal phase transitions at equilibrium, where real time becomes analogous to a control parameter such as temperature. In quenched quantum systems, recently the occurrence of DQPTs has been demonstrated, both with theory and experiment, to be intimately connected to changes of topological properties. Here, we contribute to broadening the systematic understanding of this relation between topology and DQPTs to multi-orbital and disordered systems. Specifically, we provide a detailed ergodicity analysis to derive criteria for DQPTs in all spatial dimensions, and construct basic counter-examples to the occurrence of DQPTs in multi-band topological insulator models. As a numerical case study illustrating our results, we report on microscopic simulations of the quench dynamics in the Harper-Hofstadter model. Furthermore, going gradually from multi-band to disordered systems, we approach random disorder by increasing the (super) unit cell within which random perturbations are switched on adiabatically. This leads to an intriguing order of limits problem which we address by extensive numerical calculations on quenched one-dimensional topological insulators and superconductors with disorder.
\end{abstract}

\maketitle

\section{Introduction}

Motivated by experimental progress on realizing quantum matter far from equilibrium in various physical systems including ultracold atomic gases \cite{BlochDalibardZwerger2008,GoldmanBudichZoller2016}, trapped ions \cite{BlattRoosReview2012, Jurcevic2014, Monroe2017}, nitrogen-vacancy centers in diamond \cite{Yang2019} and light-driven condensed matter systems \cite{YamamotoReview2010, Byrnes2014}, investigating the (coherent) quench dynamics of quantum many-body systems has become a broad frontier of current research \cite{PolkovnikovReview2011}. A prominent example allowing for a systematic study of intriguing non-equilibrium phenomena is provided by dynamical quantum phase transitions (DQPTs) \cite{HeylPolkovnikovKehrein2013,Karrasch2013, Canovi2014, Sharma2015, Heyl2015, VajnaDora2015, BudichHeyl2016, Zvyagin2016, Jurcevic2017, Flaschner2018, Heyl2018, Zunkovic2018, Yang2018, Qiu2018, Sedlmayr2018}, a counterpart of thermal phase transitions in coherent quantum time evolution, where the role of a control parameter is replaced by real time. 

The formal analog of a (boundary) partition function is in the context of DQPTs played by the Loschmidt amplitude
\begin{equation}
\mathcal{G}(t) = \langle \psi \vert \e^{-i H t} \vert \psi \rangle=r(t)\text{e}^{i\phi(t)},
\label{eq:LA}
\end{equation}
with $\vert \psi \rangle$ denoting the initial state and $H$ denoting the Hamiltonian governing the non-equilibrium time evolution, i.e., $\vert \psi \rangle$ is far from being an eigenstate of $H$. The role of a free energy density is assumed by the so-called rate function $g(t) = -\log(\lvert\mathcal G(t)\rvert^2) / N$, where $N$ is the size of the system, i.e., in our present context the number of lattice sites. Further following this formal analogy to thermal systems, DQPTs are then simply hallmarked by non-analytical behavior of $g$ as a function of real time, manifesting in characteristic cusps in $g(t)$ or one of its time-derivatives. These cusps are accompanied by zeros of $\mathcal{G}(t)$, known in statistical physics as Fisher zeros of the partition function \cite{Fisher1967}.

\begin{figure}[!ht]
\centering
\subfloat[geometrical phase $\phi^G(t)$]{\includegraphics[width=0.55\columnwidth]{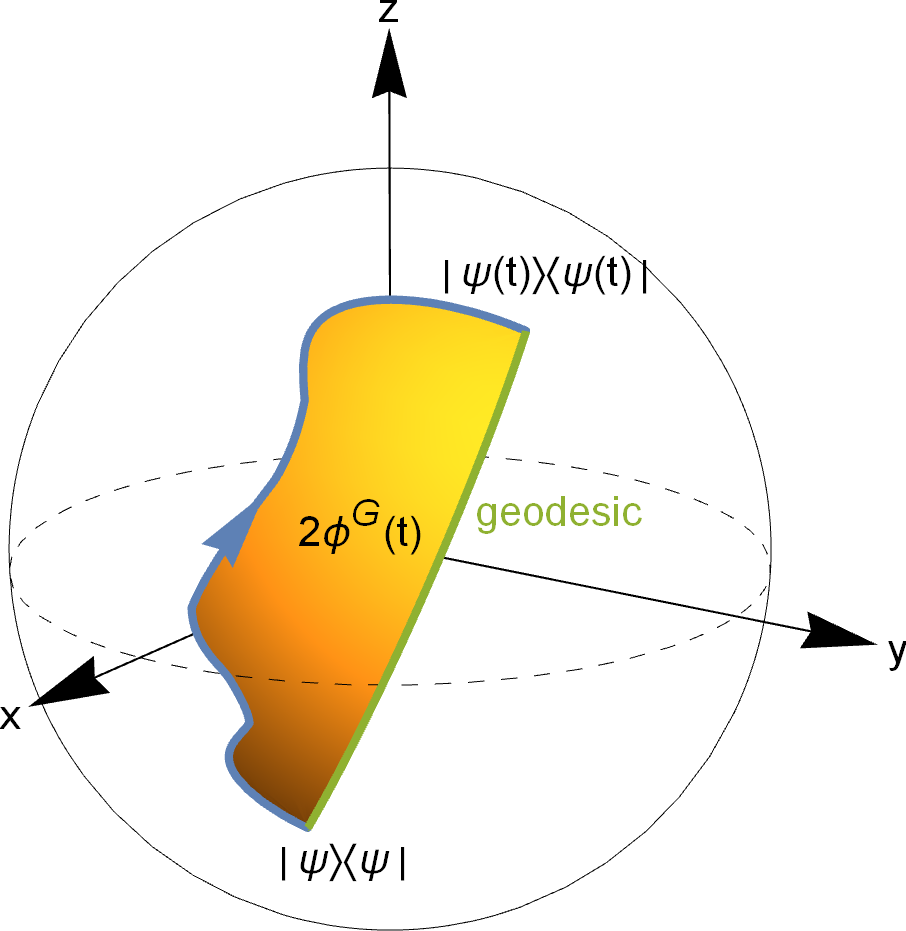} \label{fig:pancharatnam_bloch}} \\
\subfloat[$\phi^G_k(t)$ for the three-band Hofstadter model]{\includegraphics[width=\columnwidth]{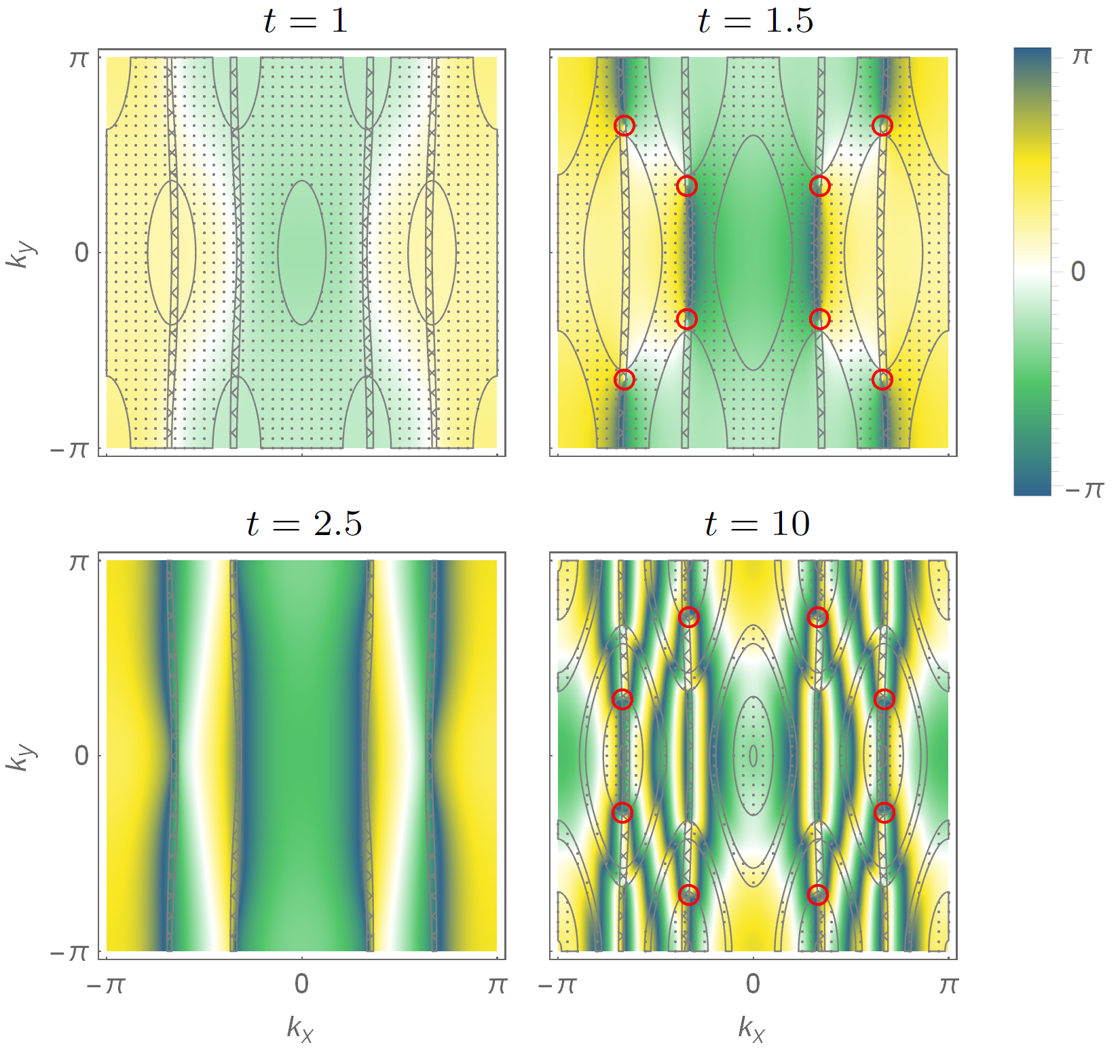} \label{fig:phiG_hofstadter_q3}}\\
\subfloat[corresponding rate function $g(t)$ and closeup of $g'(t)$]{\includegraphics[width=\columnwidth]{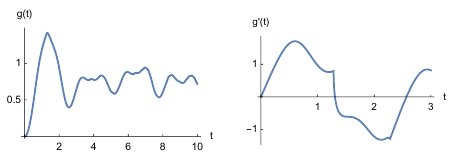} \label{fig:hofstadter_q3_rate_function}}
\caption{(a) Interpretation of the Pancharatnam geometrical phase on the Bloch sphere as half of the surface area enclosed by the trajectory up to time $t$, and the geodesic curve leading back to the initial wavefunction. (b) Time evolution snapshots of the geometrical phase for the three-band Hofstadter model after a quench. Phase vortices are circled in red, and the marked areas show the admissible region according to the criterion \eqref{eq:admissible_criterion} (cross-hatched) and the complement of the exclusion \eqref{eq:dynamical_exclusion} (dotted). (c) Corresponding rate function and its derivative. Cusps of $g'(t)$ hallmark DQPTs, i.e., the \mbox{(dis-)}appearance of Fisher zeros and phase vortex pairs.}
\label{fig:hofstadter_q3}
\end{figure}

Taking a closer look at the analytical origin of DQPTs, $\pi$-phase slips of the \emph{Pancharatnam geometrical phase} $\phi^G(t)$ \cite{Pancharatnam1956,SamuelBhandari1988} (see Fig.~\ref{fig:pancharatnam_bloch} for an illustration) have been identified as a generic phenomenon behind the non-analytical behavior of $g(t)$ \cite{BudichHeyl2016}. The phase $\phi^G(t)$ is obtained from the total phase $\phi(t)$ (see Eq.~\eqref{eq:LA}) of the complex Loschmidt amplitude by subtracting the dynamical phase
\begin{equation}
\label{eq:phiG_def}
\phi^G(t) = \phi(t) - \phi^{\text{dyn}}(t)
\end{equation}
with the dynamical phase $\phi^{\text{dyn}}(t) = - \int_0^t \ud s \, \langle \psi(s) \vert H \vert \psi(s) \rangle$. Now, when $\mathcal G(t)$ goes through a Fisher zero, its total phase $\phi(t)$ generically jumps by $\pi$, as for any zero crossing of a complex-valued function. Since the dynamical phase $\phi^{\text{dyn}}(t)$ is always continuous in time, this jump must occur in the geometrical phase $\phi^G(t)$. For the simple case of a time-dependent two-level system -- which is immediately relevant for the experimentally realized two-band models -- $\phi^G(t)$ may be readily visualized using a Bloch sphere representation (see Fig.~\ref{fig:pancharatnam_bloch}). In this picture, $\phi^G(t)$ is simply given by half of the area bounded by the time evolution trajectory between times $\tau=0$ and $\tau=t$, which is augmented to a closed path by a geodesic connecting its end points. At a Fisher zero, $\lvert\psi\rangle$ and $\lvert\psi(t)\rangle$ then correspond to antipodal points of the Bloch sphere which renders their geodesic connection (and with that $\phi^G(t)$) ill-defined. This provides a simple picture of how jumps in $\phi^G(t)$ occur at Fisher zeros hallmarking DQPTs.

Among many other intriguing applications (see Ref.~\cite{Heyl2018} for a review), DQPTs have become an important diagnostic tool for identifying topological insulator phases \cite{HasanKaneReview, QiZhangReview} in systems far from equilibrium, as has been demonstrated in recent experiments on various physical platforms, ranging from ultracold atomic gases \cite{Flaschner2018},  over superconducting qubit systems \cite{Guo2019}, and quantum walks in photonic systems \cite{Wang2019,Xu2018}, to nanomechanical settings \cite{Tian2018}. The underlying conceptual insight is that changes in the topological properties over a quench generically imply the occurrence of DQPTs \cite{VajnaDora2015,BudichHeyl2016}. Moreover, a one-to-one correspondence distinguishing such topology-induced DQPTs from accidental ones has been derived by identifying a dynamical topological order parameter for DQPTs \cite{BudichHeyl2016}. Shortly thereafter, generalizing the relation between DQPTs and topological properties, the occurrence of DQPTs in the quench dynamics in multiband topological insulators has been investigated \cite{HuangBalatsky2016}.

Our present work is aimed at further generalizing the understanding of the interplay between topology and DQPTs. More concretely, the purpose of our analysis is twofold: First, we revisit the quench dynamics in multiband systems, going beyond Ref.~\cite{HuangBalatsky2016} by providing a comprehensive ergodicity analysis resulting in criteria for DQPTs that depend on the spatial dimension of the system, and by constructing basic counter-examples to the occurrence of DQPTs in multi-band topological insulator models, where not all individual bands are topologically nontrivial (see Sec.~\ref{sec:ergodicity_analysis}). Furthermore, our results on multi-band models are supported by numerical simulations of the quench dynamics in the Hofstadter model (see Sec.~\ref{sec:Hofstadter}). Second, we connect the theory of DQPTs in multi-band and disordered systems, by approaching disorder from an angle of increasing the (super) unit cell within which random perturbations are switched on adiabatically. This leads to an intriguing order of limits problem, and to settle the question of whether topology-induced DQPTs generically survive up to a finite disorder strength, we present extensive numerical simulations on quenches in a disordered one-dimensional (1D) topological insulator model (see Sec.~\ref{sec:Disorder}).

\section{Ergodicity analysis for multiband systems}
\label{sec:ergodicity_analysis}

We consider free fermions on a (hypercubic) $d$-dimensional lattice with unit lattice constant and $n$ degrees of freedom per site. For the quantum quench, the system is assumed to be prepared in an insulating state of a filled lowest Bloch band, forming the ground state of some initial Hamiltonian $H^i$, before the system Hamiltonian is quenched at time $t=0$ to a final Hamiltonian $H$.

\subsection{Loschmidt amplitude in multiband lattice models}

Assuming lattice translation invariance, the conservation of lattice momentum allows us to factorize the Loschmidt amplitude as $\mathcal{G}(t) = \prod_{k} \mathcal{G}_k(t)$ with
\begin{equation}
\label{eq:LoschmidtG}
\mathcal{G}_k(t) = \langle \psi_{k} \rvert \e^{-i H(k) t} \lvert \psi_{k}\rangle = r_k(t) \e^{i \phi_k(t)},
\end{equation}
where $H(k)$ denotes the $n\times n$ post quench Bloch Hamiltonian in reciprocal space and $\lvert \psi_{k} \rangle$ is the occupied Bloch state of the initial Hamiltonian.

Denoting the eigenvalues and eigenvectors of the post-quench Hamiltonian $H^f(k)$ by $E_{k,\alpha}$ and $\lvert u_{k,\alpha} \rangle$, respectively, Eq.~\eqref{eq:LoschmidtG} can be written as
\begin{equation}
\mathcal{G}_k(t) = \sum_{\alpha=1}^n \left\lvert \langle u_{k,\alpha} \vert \psi_k\rangle \right\rvert^2 \e^{-i E_{k,\alpha} t}.
\label{eq:LoschmidtPoly}
\end{equation}
As mentioned, note that this formula holds for the special case of a single filled band.

\subsection{General criteria for Fisher zeros}

Because of the generalized triangle inequality in the complex plane, the occurrence of a Fisher zero at momentum $k$, i.e., $\mathcal{G}_k(t) = 0$ for some time $t$, then requires \cite{HuangBalatsky2016}
\begin{equation}
\label{eq:admissible_criterion}
\lvert \langle u_{k,\alpha} \vert \psi_k\rangle \rvert^2 \le \frac{1}{2} \quad \text{for all } \alpha=1,\ldots, n.
\end{equation}
This condition affords a simple geometric interpretation when thinking of the sum in Eq.~\eqref{eq:LoschmidtPoly} as a polygonal chain in the complex plane, the edges of which have length $\left\lvert \langle u_{k,\alpha} \vert \psi_k\rangle \right\rvert^2$ that rotate with independent frequencies $E_{k,\alpha}$: A violation of Eq.~\eqref{eq:admissible_criterion} then simply means that one edge dominates in length over all others such that concatenating all edges can never lead to a closed polygon, independent of their direction.

Another relevant criterion for the (non-)occurrence of Fisher zeros at a fixed time $t$ is whether the points $\{\e^{-i E_{k,\alpha} t}\}_{\alpha = 1, \dots, n}$, all lie within a minor arc of the unit circle; equivalently, whether the convex polygon with vertices $\{\e^{-i E_{k,\alpha} t}\}_{\alpha = 1, \dots, n}$ (as points in the complex plane) contains the origin. In other words, if there exists a $\omega(t) \in \R$ such that
\begin{equation}
\label{eq:dynamical_exclusion}
\cos\!\left(E_{k,\alpha} t - \omega(t)\right) > 0 \quad \text{for all } \alpha=1,\ldots, n,
\end{equation}
then the sum in Eq.~\eqref{eq:LoschmidtPoly} cannot be zero.

Note that the condition \eqref{eq:admissible_criterion} only depends on the initial state and the eigenvectors of $H^f(k)$, whereas the dynamical criterion \eqref{eq:dynamical_exclusion} solely depends on the eigenvalues of $H^f(k)$ and time $t$.

\subsection{Abundance of Fisher zeros}
\label{sec:abundance_fisher_zeros}

In Ref.~\cite{HuangBalatsky2016}, it has been shown that quenches from a trivial initial state into a post-quench Hamiltonian, all individual bands of which have non-zero Chern number, there must be a momentum for which Eq.~\eqref{eq:admissible_criterion} is satisfied. Basic ergodicity arguments then imply that $\mathcal{G}_k(t)$ must come arbitrarily close to zero at some finite time $t$. However, these important insights do not yet provide a sufficient condition for the actual occurrence of a Fisher-zero, i.e., an exact zero crossing of $\mathcal{G}_k(t)$ at any finite time. In the following, we fill this gap by performing an additional dimensional analysis, revealing also the generic dependence of the abundance of Fisher zeros on the spatial dimension $d$. We note that zeros of the partition function in the complex plane have been studied in the context of phase transitions for more than 50 years, including the analysis of the dimensional dependence of critical exponents \cite{Fisher1967, Abe1967b, Abe1967c, Suzuki1967, Grossmann1969a, Grossmann1969b, Saarloos1984}.

We start by observing that Eq.~\eqref{eq:admissible_criterion} for $n>2$ is generically satisfied in an entire {\textit{admissible region}} of spatial dimension $d$, i.e., in a whole neighborhood in momentum space. Therefore, as a subset of the $(d+1)$-dimensional momentum-time space (where momentum space is constrained to the admissible region), the dimension of the manifold of Fisher zeros $\mathcal{G}_k(t) = 0$ is generically given by $(d + 1) - 2 = d - 1$, since both the real and imaginary parts of $\mathcal{G}_k(t)$ have to be tuned to zero. This dimensional counting is independent of $n$ for $n>2$, again since $(\e^{-i E_{k,1} t},\ldots,\e^{-i E_{k,n} t})$ is ergodic on the $n$-dimensional torus as long as the energies are rationally independent. As a consequence, in a one-dimensional system ($d=1$), the Fisher zeros are expected to occur at isolated points in time-momentum space, while for $d=2$, the set of Fisher zeros are curves in the three-dimensional momentum-time space, in agreement with microscopic simulations on the quench-dynamics of two-band models in $d=2$ \cite{HeylBudich2017}.

We now elaborate on the somewhat exceptional but experimentally highly relevant case $n=2$. There, Eq.~\eqref{eq:admissible_criterion} implies $\lvert \langle u_{k,1} \vert \psi_k\rangle \rvert^2=\lvert \langle u_{k,2} \vert \psi_k\rangle \rvert^2=1/2$ which generically is only satisfied in a $(d-1)$-dimensional admissible region, rather than the $d$-dimensional neighborhood found for $n>2$. However, this reduction in dimension of the set of admissible momenta for $n=2$ is exactly compensated by the fact that then $\mathcal{G}_k(t) = \e^{-it( E_{k,1}+E_{k,2})/2} \cos(t( E_{k,1}-E_{k,2})/2)$ in the admissible region which requires only tuning of a single real condition (the argument of the $\cos$-function) in order to achieve zeros. Hence, Fisher zeros are now guaranteed to occur at all admissible momenta, namely at the times $t_{k,l}=(2l+1)\pi/(E_{k,1}-E_{k,2}),~l=1,2,\ldots$ such that they after all still form a $(d-1)$-dimensional set, similar to the $n>2$ case.

\subsection{Avoided DQPTs in quenched Chern insulators}

Quenches from trivial states to Chern insulator Hamiltonians imply DQPTs, at least when assuming that all individual bands of the post quench Hamiltonian have non-zero Chern number \cite{HuangBalatsky2016}. To demonstrate that this quite strong assumption is indeed necessary, we construct a basic counter-example, where the post quench Hamiltonian is in a Chern insulator regime, but where no Fisher zeros or DQPTs occur as not all individual bands have non-vanishing Chern number. To this end, consider a system with three bands, where we quench from an initial Hamiltonian with only topologically trivial bands to a Chern insulator which has Chern numbers $(1,0,-1)$, ordered from the lowest to the highest band. Now we assume that the lowest band of the initial Hamiltonian is formed by Bloch functions that have a large overlap ($>1/\sqrt{2}$ at all momenta) with the trivial central band of the post quench Hamiltonian. In this case, the Fisher-zero admissibility criterion \eqref{eq:admissible_criterion} can never be satisfied, and, as a consequence, no Fisher zeros or DQOTs occur at any time.

\section{Quenched Hofstadter model}
\label{sec:Hofstadter}

In this section, we practically verify our general ergodicity analysis by time-dependent simulations of DQPTs in multi-band systems. For concreteness, we consider the $q$-band (magnetic flux $2\pi/q$ per unit cell) Hofstadter model \cite{Hofstadter1976,Aidelsburger2013} defined in the Landau gauge by the momentum space (Bloch) Hamiltonian
\begin{equation}
\label{eq:Hhofstadter}
\begin{split}
&H(k) = \\
&\begin{pmatrix}
2 \cos(k_x) & 1                             &        &    \e^{-i k_y}                      \\
1           & 2 \cos(k_x - \tfrac{2\pi}{q}) & 1      &                                     \\
            &                               & \ddots &                                     \\
\e^{i k_y}  &                               & 1      & 2 \cos(k_x - \tfrac{2\pi (q-1)}{q}) \\
\end{pmatrix}.
\end{split}
\end{equation}

Because of conservation of lattice momentum, the Loschmidt amplitude factorizes (see Eq.~\eqref{eq:LoschmidtG}). We consider two scenarios for the topologically trivial initial state $\vert \psi_k\rangle$: (i) the initial state occupies the first orbital (in the basis of Eq.~(\ref{eq:Hhofstadter})) for each $k$, i.e., $\lvert\psi_k\rangle = \lvert e_1 \rangle$, and (ii) $\lvert\psi_k\rangle$ is equal to a fixed complex-random state (independent of $k$). The second scenario will exemplify the absence of symmetries beyond lattice momentum conservation.

Fig.~\ref{fig:phiG_hofstadter_q3} shows snapshots of $\phi^G_k(t)$ at several points in time, for scenario (i) and $q = 3$. Fisher zeros, at which $\phi^G_k(t)$ is ill-defined, appear as phase vortices at isolated $k$-points (circled in red) which contain the whole range of phases, $[-\pi, \pi]$, in any (arbitrarily small) neighborhood. This is in line with the dimension analysis in Sec.~\ref{sec:abundance_fisher_zeros}: the Fisher zeros should describe a $d - 1 = 1$ dimensional submanifold within momentum-time space. Concretely, within a certain interval of time, Fisher zeros are found at all points in time at isolated momenta.

The cross-hatched areas in Fig.~\ref{fig:phiG_hofstadter_q3} show the static (time-independent) admissible region defined via Eq.~\eqref{eq:admissible_criterion}, and the dotted areas the complement of the dynamical exclusion criterion \eqref{eq:dynamical_exclusion}, solely depending on the eigenvalues and $t$. Indeed the phase vortices stay inside both regions, as required. Note that the stripe-shaped pattern of the static admissible region (for the present model parameters) implies that phase vortex--antivortex pairs are constrained to remain within a single stripe. At $t = 2.5$, the dynamical exclusion holds within the entire Brillouin zone, thus disallowing any Fisher zeros. We note that in general the admissible regions have a different dimension as compared to the manifold of Fisher zeros.

Fig.~\ref{fig:hofstadter_q3_rate_function} visualizes the corresponding rate function $g(t)$ and a closeup of its derivative. Because of the factorization $\mathcal{G}(t) = \prod_{k} \mathcal{G}_k(t)$, the rate function equals
\begin{equation}
g(t) = - \frac{1}{N} \log\left(\lvert\mathcal G(t)\rvert^2\right) = - \frac{1}{\lvert\text{BZ}\rvert} \int_{\text{BZ}} \ud k \log\left(\lvert\mathcal G_k(t)\rvert^2\right)
\end{equation}
with $\text{BZ} = [-\pi, \pi]^2$ being the Brillouin zone of the present model. The (weak) log-singularity of the integrand at Fisher zeros leads to a cusp in the derivative $g'(t)$ at their \mbox{(dis-)}appearance, as visible in Fig.~\ref{fig:hofstadter_q3_rate_function}. Specifically, Fisher zeros occur for the first time around $t = 1.3$ (first cusp) and then disappear around $t = 2.3$ (second cusp).

To systematically understand the symmetries of the phase pattern of $\phi^G_k(t)$  with respect to lattice momentum, first note that $H(k_x, -k_y) = H(k)^T$ according to Eq.~\eqref{eq:Hhofstadter}, such that for \emph{real-valued} $\lvert\psi_k\rangle$, $\mathcal{G}_{(k_x,-k_y)}(t) = \mathcal{G}_{k}(t)$. In particular, this mirror symmetry holds in the first scenario. Moreover, central inversion ($k \to -k$) can be expressed as unitary transformation: Let $P_q$ be the $q \times q$ permutation matrix which sends the $j$-th entry of a vector (counting from zero) to $-j \mod q$ ($j = 0, \dots, q-1$), and define
\begin{equation}
U(k_y) = P_q \cdot \begin{pmatrix} \e^{-i k_y} & & & \\ & 1 & & \\ & & \ddots & \\ & & & 1 \end{pmatrix}.
\end{equation}
Then
\begin{equation}
U(k_y)^{\dagger} H(k) U(k_y) = H(-k).
\end{equation}
It follows that
\begin{equation}
\mathcal{G}_{-k}(t) = \langle U(k_y)\psi_{-k} \rvert \e^{-i H(k) t} \lvert U(k_y)\psi_{-k}\rangle.
\end{equation}
Since $U(k_y) \lvert e_1 \rangle = \e^{-i k_y} \lvert e_1 \rangle$ and since the phase factor $\e^{-i k_y}$ cancels in $\mathcal{G}_{-k}(t)$, this explains the inversion symmetry apparent in Fig.~\ref{fig:phiG_hofstadter_q3}.

\begin{figure}[!ht]
\centering
\subfloat[$\phi^G_k(t)$ for the Hofstadter model with random initial state]{\includegraphics[width=\columnwidth]{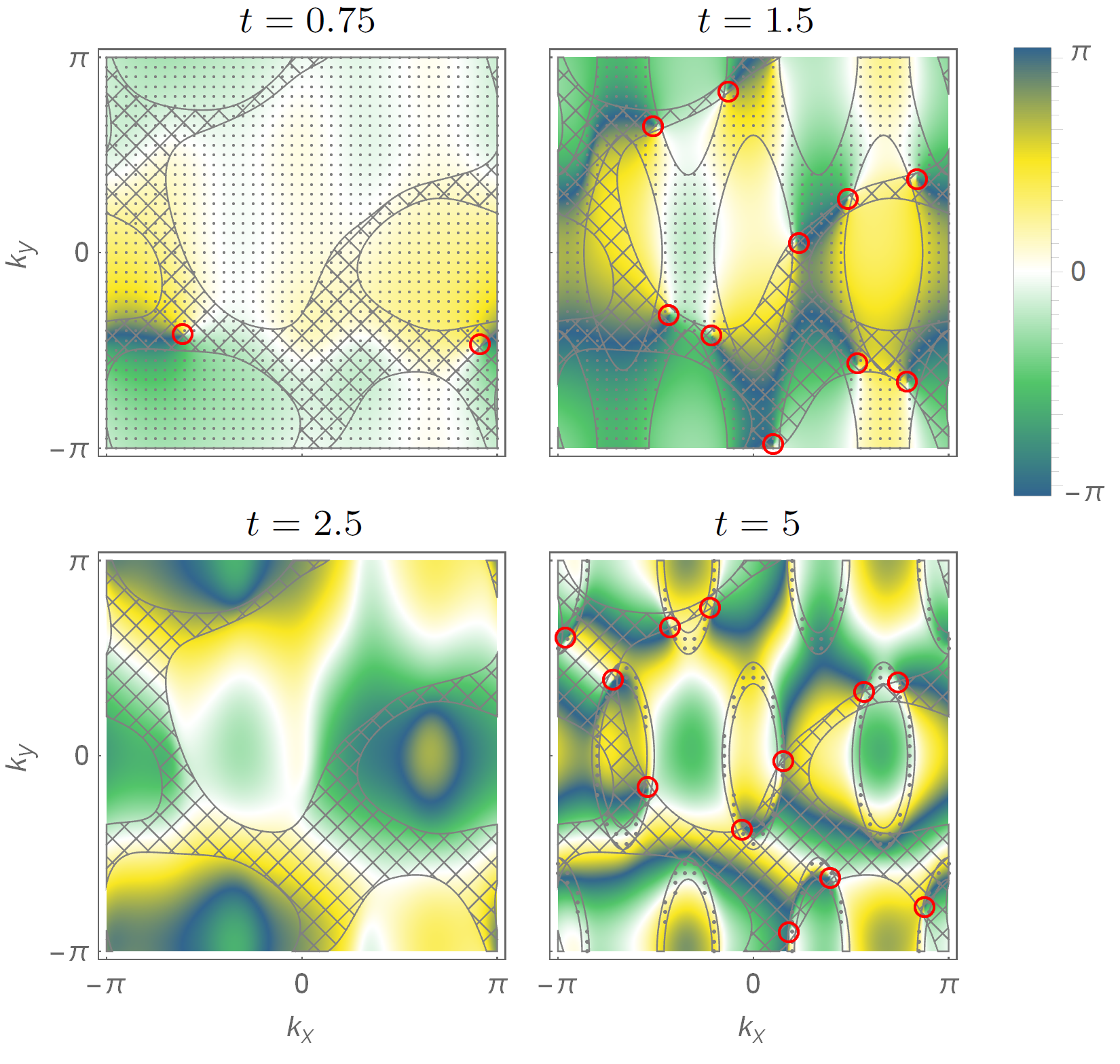}}\\
\subfloat[corresponding rate function $g(t)$ and closeup of $g'(t)$]{\includegraphics[width=\columnwidth]{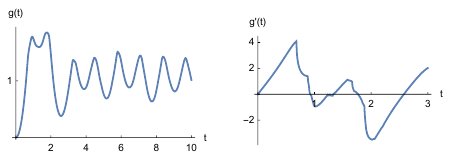}}
\caption{Pancharatnam geometrical phase and rate function for the three-band Hofstadter model as in Fig.~\ref{fig:hofstadter_q3}, but for a ($k$-independent) initial state with complex random entries. The lack of momentum symmetry of the geometrical phase is expected (see main text). Note that the dynamical criterion for Fisher zeros (dotted areas) only depends on the spectrum of $H(k)$ and thus agrees with Fig.~\ref{fig:phiG_hofstadter_q3}.}
\label{fig:hofstadter_q3_psi0rnd}
\end{figure}

In contrast, for the second scenario (ii) of a complex random initial state, our analysis does not predict any momentum symmetry. Fig.~\ref{fig:hofstadter_q3_psi0rnd} shows the geometrical phase and rate function for the second scenario, and indeed momentum symmetry is now absent. Nevertheless, the dynamical exclusion criterion in Eq.~\eqref{eq:dynamical_exclusion} only depends on the spectrum of $H(k)$ and time, and thus agrees for both scenarios. In particular, it disallows any Fisher zeros at $t = 2.5$, as in the first scenario.

\section{Disordered systems}
\label{sec:Disorder}

\begin{figure*}
\centering
\includegraphics[width=\textwidth]{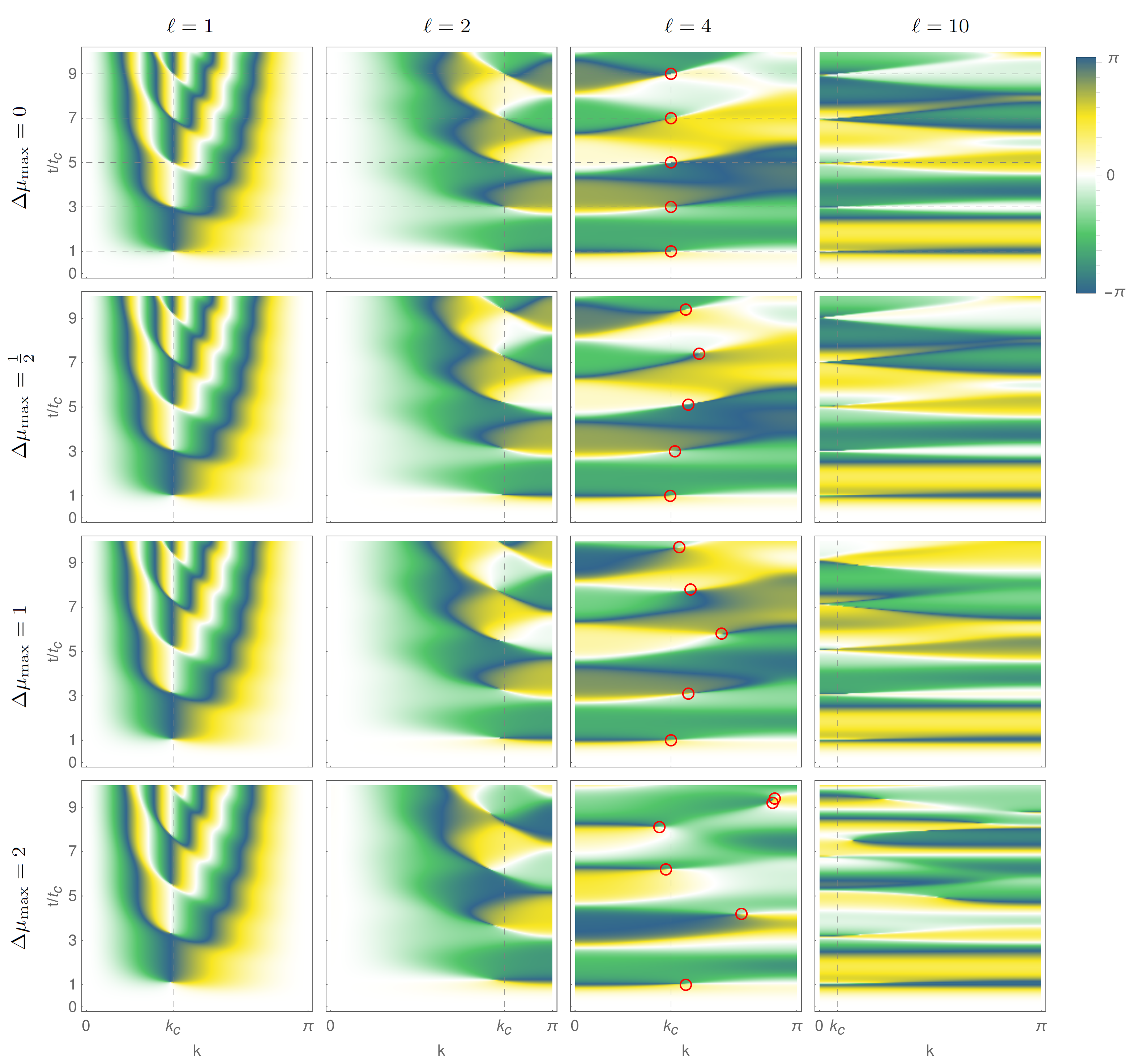}
\caption{Pancharatnam geometrical phase for the disordered Kitaev chain with period $\ell$ and increasing disorder strength $\Delta\mu_{\text{max}}$. Each row corresponds to a fixed disorder strength (starting from zero disorder in the top row), and each column to a fixed supercell size $\ell$. The dashed vertical lines mark the critical momentum $k_c$ of the ordered system ($\Delta\mu_{\text{max}} = 0$).}
\label{fig:phiG_kitaev}
\end{figure*}

\subsection{General framework}

We now gradually extend the framework of  Pancharatnam geometric phase vortices leading to DQPTs from multi-orbital to disordered systems. To this end we consider systems that are still periodic, but with respect to a super cell containing $\ell \gg 1$ lattice sites. Within the supercell, disorder is modeled by adding random, spatially uncorrelated perturbations to the Hamiltonian coefficients in a real-space representation, concretely by changing the onsite potential term $\mu \, \cdag_j \cann_j$ to $\mu_j\,\cdag_j \cann_j$ with $\mu_j = \bar{\mu} + \Delta \mu_j$ and $\Delta \mu_j$ the perturbation. For sufficiently large  $\ell$, the system resembles a disordered system (without any periodicity), as the relevant physical properties are expected to be negligibly changed when matching distant coefficients, i.e., $\Delta \mu_{j + \ell} = \Delta \mu_j$. The momentum representation of the Hamiltonian is now based on a supercell of size $\ell$. For example, an unperturbed Hamiltonian in Bogoliubov-de Gennes form
\begin{equation}
H = \frac{1}{2\pi} \int_{\mathbb{T}} \ud k \, \begin{pmatrix} \chatdag_k & \chatann_{-k} \end{pmatrix} \left(\vec{d}(k) \cdot \vec{\sigma}\right) \begin{pmatrix} \chatann_k \\ \chatdag_{-k} \end{pmatrix}
\end{equation}
(with $\vec{\sigma}$ the vector of Pauli matrices) is changed to
\begin{equation}
H^{\ell} = \frac{1}{2\pi} \int_{\mathbb{T}} \ud k \, (\hat{\chi}^{\ell}_k)^{\dagger} \, h^{\ell}(k) \, \hat{\chi}^{\ell}_k
\end{equation}
with
\begin{equation}
\label{eq:chi_ell_k_def}
\hat{\chi}^{\ell}_k = \begin{pmatrix}
\chatann_{k,0} & \chatdag_{-k,0} & \cdots & \chatann_{k,\ell-1} & \chatdag_{-k,\ell-1} \end{pmatrix}^T
\end{equation}
and $h^{\ell}(k)$ being a $2\ell \times 2\ell$ matrix depending on the disorder realization. The index $\alpha$ in $\chatann_{k,\alpha}$ appearing in \eqref{eq:chi_ell_k_def} may be interpreted as orbital index.

The Loschmidt amplitude defined in \eqref{eq:LoschmidtG} becomes in the supercell representation
\begin{equation}
\label{eq:LoschmidtG_disorder}
\mathcal{G}^{\ell}_k(t) = \det\left( \big\langle \psi_{k,j} \vert \e^{-i h^{\ell}(k) t} \vert \psi_{k,j'} \big\rangle \right)_{j,j'=1}^{\ell}
\end{equation}
with the orthonormal $\psi_{k,j}$, $j = 1, \dots, \ell$ defining the initial state as Slater determinant $\lvert \psi_{k,1} \cdots \psi_{k,\ell} \rangle$ of occupied modes. We denote the complex phase of $\mathcal{G}^{\ell}_k(t)$ by $\phi^{\ell}_k(t)$. Note that one recovers the special case of zero noise as
\begin{equation}
\label{eq:LoschmidtG_zero_disorder}
\mathcal{G}^{\ell}_k(t) = \prod_{j=1}^{\ell} \big\langle \psi_{k,j} \vert \e^{-i h^{\ell}(k) t} \vert \psi_{k,j} \big\rangle
\end{equation}
since the matrix in \eqref{eq:LoschmidtG_disorder} can then be canonically diagonalized due to translation invariance. In particular, the corresponding phase $\phi^{\ell}_k(t)$ is then given by the following $\ell$-fold superposition of phases:
\begin{equation}
\label{eq:phase_superposition}
\phi^{\ell}_k(t) = \sum_{j=1}^{\ell} \phi_{k,j}(t) \mod 2\pi.
\end{equation}
The dynamical phase reads in the supercell representation
\begin{equation}
\phi^{\text{dyn},\ell}_k(t) = - t \sum_{j=1}^{\ell} \langle \psi_{k,j} \vert h^{\ell}(k) \vert \psi_{k,j} \rangle \mod 2\pi,
\end{equation}
and analogously $\phi^{G,\ell}_k(t) = \phi^{\ell}_k(t) - \phi^{\text{dyn},\ell}_k(t)$.

Since $\mathcal{G}^{\ell}_k(t)$ is a real-analytic function of the noise coefficients (such as $\Delta \mu_j$ in the example), the non-analytic points of the Pancharatnam geometrical phase (i.e., Fisher zeros of the Loschmidt amplitude) cannot instantaneously disappear when continuously increasing the noise strength; instead, the non-analytic points will continuously move in the $k$-$t$-plane, potentially annihilating or being created in pairs.

\subsection{Disordered Kitaev chain}
\label{sec:disordered_kitaev_chain}

As a specific example, we investigate the Kitaev chain \cite{Kitaev2001,Beenakker2013} described by the Hamiltonian
\begin{multline}
\label{eq:HKitaev}
H = \sum_{j \in \Z} \Big[ -t \left(\cdag_j \cann_{j+1} + \text{h.c.}\right) + \mu \left(\cdag_j \cann_j - \tfrac{1}{2}\right) \\
+ \left(\Delta \cann_j \cann_{j+1} + \text{h.c.}\right) \Big]
\end{multline}
where $t$ is the hopping amplitude, $\mu$ is the chemical potential and $\Delta$ is the superconducting gap.

Switching to the Bogoliubov-de Gennes momentum representation of the Hamiltonian,
\begin{equation}
H(k) = \vec{d}(k) \cdot \vec{\tau}
\end{equation}
with $\vec{d}(k) = \left( 0, \Delta \sin(k), \frac{\mu}{2} - t \cos(k) \right)$ and $\vec{\tau}$ the Nambu pseudospin, one obtains the Pancharatnam geometrical phase defined in \eqref{eq:phiG_def}, which allows to identify singular points of the Loschmidt amplitude \cite{BudichHeyl2016}. We now employ the supercell representation to investigate the effects of disorder (see also Appendix~\ref{sec:GeneralKitaevChain} for technical details): For simplicity, we solely let the chemical potential in \eqref{eq:HKitaev} be site-dependent, i.e., $\mu_j = \bar{\mu} + \Delta \mu_j$ with independent and identically distributed random variables $\Delta \mu_j$ chosen from some interval $[-\Delta \mu_{\text{max}}, \Delta \mu_{\text{max}}]$ (uniformly distributed); we retain periodicity with period $\ell \in \N$, i.e.,\ $\mu_{j + \ell} = \mu_j$ for all $j \in \Z$. The specific parameters for the following are $t = 1$, $\bar{\mu} = 6$ and $\Delta = 1$. We checked, however, that different disorder scenarios, such as adding noise to the hopping amplitudes or superconducting gap parameters instead of the potentials, lead to qualitatively similar findings regarding the physics of DQPTs. In particular, the cusps of the rate function (see below) remain intact. This holds even though these scenarios differ regarding their effectiveness in localizing the eigenstates of the Hamiltonian. 

Fig.~\ref{fig:phiG_kitaev} shows the Pancharatnam geometrical phase for a fixed noise realization but increasing noise strength, and various supercell sizes $\ell$. According to Eq.~\eqref{eq:phase_superposition}, the supercell representation effectively folds back the phase along the momentum direction. Accordingly, the geometrical phase assumes a stripe-like pattern with increasing $\ell$, i.e., it varies less as a function of momentum.

Using the supercell representation, the rate function for the present model reads
\begin{equation}
\label{eq:rate_function_supercell_def}
g^{\ell}(t) = - \frac{1}{\pi} \int_0^\pi \ud k \log\!\left[ \big\lvert \mathcal{G}^{\ell}_k(t) \big\rvert \right] / \ell.
\end{equation}
Thus the zeros of $\mathcal{G}^{\ell}_k(t)$ result in (weak) log-singularities of the integrand and corresponding cusps of $g^{\ell}(t)$.

\begin{figure}
\centering
\subfloat[$g^{\ell}(t)$ and $\frac{\ud}{\ud t} g^{\ell}(t)$ around $t_c$ for $\ell = 4$]{\includegraphics[width=\columnwidth]{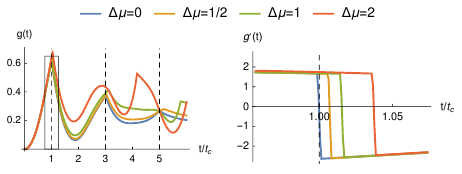}} \\
\subfloat[$g^{\ell}(t)$ and $\frac{\ud}{\ud t} g^{\ell}(t)$ around $t_c$ for $\Delta\mu_{\text{max}} = \frac{1}{2}$]{\includegraphics[width=\columnwidth]{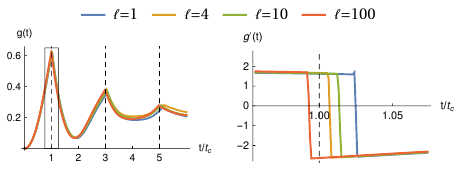}}
\caption{Rate function and its derivative around the first critical time point $t_c$, for the disordered Kitaev chain with random disorder realizations in the supercell representation.}
\label{fig:kitaev_chain_disorder_rate_function}
\end{figure}

Fig.~\ref{fig:kitaev_chain_disorder_rate_function} visualizes $g^{\ell}(t)$ for the disordered Kitaev chain and random disorder realizations, illustrating (a) the effect of increasing disorder at fixed supercell size and (b) increasing supercell size at fixed disorder strength. The critical time $t_c$ of the first Fisher zero for the case without disorder has been obtained semi-analytically \cite{BudichHeyl2016}. One observes in Fig.~\ref{fig:kitaev_chain_disorder_rate_function}a that the rate function is continuously deformed with increasing noise strength, and while the time points of the cusps (i.e., Fisher zeros) shift, the cusps do not instantaneously disappear (see also the time derivative around $t_c$ on the right). This is expected due to the real-analytic dependence of the Loschmidt amplitude on the noise coefficients, as detailed above. Visually, the perseverance of the cusps can be understood based on the geometrical phase in Fig.~\ref{fig:phiG_kitaev}. Namely, the cusps correspond precisely to the phase vortices, and thus the \mbox{(dis-)}appearance of cusps and vortex--antivortex pairs at momenta $k$ and $-k$ with increasing disorder strength is equivalent. This does not happen instantaneously when turning on disorder at finite $\ell$, since the vortex positions depend continuously on the disorder strength and have a finite distance in momentum at zero disorder.

However, in the limit $\ell \rightarrow \infty$ the size of the effective Brillouin zone associated with the supercell shrinks to zero, leading to a non-trivial order of limits problem for the stability of DQPTs against disorder. To settle this issue, we performed extensive numerical simulations on systems with finite disorder strength and large $\ell$. Our results, summarized in Fig.~\ref{fig:kitaev_chain_disorder_rate_function}, give strong numerical evidence that the non-analyticities in the rate function hallmarking DQPTs persist up to significant disorder strength even in the large $\ell$ limit, i.e., when approaching the disordered case without residual translational invariance. 

\begin{figure}
\centering
\subfloat[histogram]{\includegraphics[width=0.8\columnwidth]{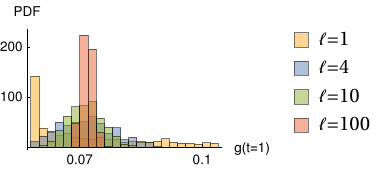}} \\
\subfloat[variance]{\includegraphics[width=0.5\columnwidth]{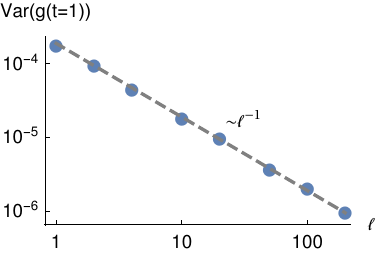}}
\caption{Histogram and corresponding variance for random disorder realizations ($\Delta\mu_{\text{max}} = \frac{1}{2}$) of the rate function $g^{\ell}(t)$ evaluated at $t = 1$. The variance exhibits a $\sim 1/\ell$ scaling, with $\ell$ the supercell size.}
\label{fig:kitaev_chain_disorder_rate_function_statistics}
\end{figure}

Having investigated instances of disorder realizations so far, we will now analyze averaging effects as the supercell size increases. Fig.~\ref{fig:kitaev_chain_disorder_rate_function_statistics} shows the histogram and corresponding variance of the rate function $g^{\ell}(t)$ evaluated at time point $t = 1$, for various supercell sizes $\ell$. The observed $\sim 1/\ell$ scaling of the variance is likely due to the disorder contributions from individual lattices sites being (almost) independent, analogous to the sum of independent random variables in the central limit theorem. This becomes plausible when assuming that \eqref{eq:LoschmidtG_zero_disorder} holds approximately for weak disorder, and by inserting \eqref{eq:LoschmidtG_zero_disorder} into \eqref{eq:rate_function_supercell_def} we get:
\begin{equation}
g^{\ell}(t) \approx - \frac{1}{\ell} \sum_{j=1}^{\ell} \frac{1}{\pi} \int_0^\pi \ud k \log\!\left[ \big\lvert \big\langle \psi_{k,j} \vert \e^{-i h^{\ell}(k) t} \vert \psi_{k,j} \big\rangle \big\rvert \right].
\end{equation}
Now a lattice to momentum transformation applied to $h^{\ell}(k)$ may be understood as an orthogonal transformation of the random disorder coefficients, and if these are multivariate normal distributed, the transformed coefficients will remain independent.

\section{Concluding discussion}

We investigated the stability and topological properties of dynamical quantum phase transitions going beyond the minimal setting of lattice translation invariant two-band models in two somewhat related directions. First, building up on recent results \cite{HuangBalatsky2016} on the occurrence of DQPTs in multi-orbital systems, we demonstrated how the phenomenology of DQPTs depends on the spatial dimension of the system by means of a more in depth ergodicity analysis of the Loschmidt amplitude. We emphasize that our analysis (and Ref.~\cite{HuangBalatsky2016}) was based on the assumption of a single filled band. Hence, the derivation of strict criteria for the occurrence of DQPTs in multi-band systems with more than one occupied bands remains an interesting subject of future research. 

Second, we considered random potential fluctuations within a (super) unit cell of increasing size as a route towards understanding the stability of DQPTs in disordered systems. This approach yielded clear analytical insights supporting for the considered settings the stability of DQPTs for finite unit cells with random potential. However, a non-trivial order of limits problem renders an analytical proof for the truly disordered case of an infinite spatial period of the random potential elusive. To fill this gap, at least for the considered model systems, we presented numerical simulations for systems with large unit cells, thus corroborating the existence of DQPTs as hallmarked by non-analyticities of the rate function up to significant disorder strength.

The numerical simulations presented in this work encourage accompanying theoretical investigations: Specifically, a promising direction could be a perturbation analysis (with respect to disorder strength) applied to Eq.~\eqref{eq:LoschmidtG_disorder}, which should result in \eqref{eq:LoschmidtG_zero_disorder} as a lowest order term. Also, the question of whether and to what extent disorder contributions to the rate function can indeed be treated as independent (as conjectured in Sec.~\ref{sec:disordered_kitaev_chain}) may be settled in future work.

We close by briefly discussing the relation of our present analysis to recent other studies on the combination of disorder and DQPTs. In Ref.~\cite{Yang2017}, the interplay between quasi-periodic potentials and DQPTs has been investigated, demonstrating the existence of Fisher zeros in certain limits of quasi-disorder, and identifying the value of the Loschmidt echo as a marker for localization. Shortly after, in Ref. \cite{Yin2018}, DQPTs have been exemplified to serve as a tool for diagnosing Anderson localization transitions in certain disordered 1D and 3D models. Very recently, the effect of disorder on DQPTs in extended toric code models has been analyzed \cite{Srivastav2018}. Approaching the fate of DQPTs in disordered systems by following vortices in the geometric phase in systems with a growing disordered super-cell, however, is unique to our present work.

\begin{acknowledgments}
We acknowledge helpful discussions with Markus Heyl. J.C.B.\ acknowledges financial support from the German Research Foundation (DFG) through the Collaborative Research Centre SFB 1143 (Project No.~247310070) and the W\"urzburg-Dresden Cluster of Excellence on Complexity and Topology in Quantum Matter -- ct.qmat (EXC 2147, Project No.~39085490).
\end{acknowledgments}

\pagebreak

\appendix

\section{Generalized Kitaev chain with periodic supercell structure}
\label{sec:GeneralKitaevChain}

We consider the Kitaev chain as in \eqref{eq:HKitaev}, generalized to site-dependent coefficients, i.e.,
\begin{multline}
\label{eq:generalHKitaev}
H = \sum_{j \in \Z} \Big[ -t_j \left(\cdag_j \cann_{j+1} + \text{h.c.}\right) + \mu_j \left( \cdag_j \cann_j - \tfrac{1}{2} \right) \\
+ \left(\Delta_j \cann_j \cann_{j+1} + \text{h.c.}\right) \Big].
\end{multline}

The Hamiltonian may formally be represented in Bogoliubov-de Gennes form as
\begin{multline*}
H = \begin{pmatrix} \cdots & \cdag_j & \cann_j & \cdag_{j+1} & \cann_{j+1} & \cdots\end{pmatrix} \\
\times \begin{pmatrix}
\ddots & \ddots            & \ddots          &         &         &        \\
       & B_{j-1}^{\dagger} & A_{j}           & B_{j}   &         &        \\
       &                   & B_{j}^{\dagger} & A_{j+1} & B_{j+1} &        \\
       &                   &                 & \ddots  & \ddots  & \ddots \\
\end{pmatrix}
\begin{pmatrix}
\vdots \\ \cann_j \\[0.5em] \cdag_j \\[0.5em] \cann_{j+1} \\[0.5em] \cdag_{j+1} \\ \vdots
\end{pmatrix}
\end{multline*}
with $2 \times 2$ blocks
\begin{equation}
A_j = \frac{1}{2} \begin{pmatrix} \mu_j & 0 \\ 0 & -\mu_j \end{pmatrix} \quad \text{and} \quad
B_j = \frac{1}{2} \begin{pmatrix} -t_j & -\Delta_j^* \\ \Delta_j & t_j \end{pmatrix}.
\end{equation}

In the following, we assume periodicity with period $\ell \in \N$, i.e.,\ $t_{j + \ell} = t_j$, $\mu_{j + \ell} = \mu_j$ and $\Delta_{j + \ell} = \Delta_j$ for all $j \in \Z$. Thus we may subsume the creation and annihilation operators in a spinor
\begin{equation}
\chi^{\ell}_n = \begin{pmatrix}
\cann_{\ell n} & \cdag_{\ell n} & \cdots & \cann_{\ell n + \ell - 1} & \cdag_{\ell n + \ell - 1} \end{pmatrix}^T
\end{equation}
and represent the Hamiltonian as
\begin{equation}
\label{eq:HKitaev_supercell}
H = \sum_{n \in \Z} \left[ (\chi^{\ell}_n)^{\dagger} \, h^{\ell}_{\text{local}} \, \chi^{\ell}_n + \left((\chi^{\ell}_n)^{\dagger} \, h^{\ell}_{\text{hop}} \, \chi^{\ell}_{n+1} + \text{h.c.}\right)\right]
\end{equation}
with
\begin{equation}
h^{\ell}_{\text{local}} =
\begin{pmatrix}
A_0           & B_0    &                   &         \\
B_0^{\dagger} & A_1    & B_1               &         \\
              & \ddots & \ddots            & \ddots  \\
              &        & B_{\ell-2}^{\dagger} & A_{\ell-1} \\
\end{pmatrix}
\end{equation}
and
\begin{equation}
h^{\ell}_{\text{hop}} =
\begin{pmatrix}
0          &   &        & 0 \\
           & 0 &        &   \\
           &   & \ddots &   \\
B_{\ell-1} &   &        & 0 \\
\end{pmatrix}.
\end{equation}

To arrive at a momentum representation of the Hamiltonian, we use Fourier transformation
\begin{equation}
\label{eq:spinor_fourier}
\chi^{\ell}_n = \frac{1}{2\pi} \int_{\mathbb{T}} \ud k \e^{i k n} \hat{\chi}^{\ell}_k
\end{equation}
with
\begin{equation}
\hat{\chi}^{\ell}_k = \begin{pmatrix}
\chatann_{k,0} & \chatdag_{-k,0} & \cdots & \chatann_{k,\ell-1} & \chatdag_{-k,\ell-1} \end{pmatrix}^T.
\end{equation}
Here the index $\alpha$ in $\chatann_{k,\alpha}$ may be interpreted as orbital index. The first Brillouin zone is equal to the interval $\mathbb{T} = [-\pi, \pi]$ with periodic boundary conditions. Inserting \eqref{eq:spinor_fourier} into \eqref{eq:HKitaev_supercell} yields
\begin{equation}
\label{eq:HKitaev_supercell_k}
H = \frac{1}{2\pi} \int_{\mathbb{T}} \ud k \, (\hat{\chi}^{\ell}_k)^{\dagger} \left[ h^{\ell}_{\text{local}} + \left(\e^{i k} h^{\ell}_{\text{hop}} + \text{h.c.}\right)\right] \hat{\chi}^{\ell}_k.
\end{equation}

Note that the conventional momentum representation of the Kitaev chain is recovered for $\ell = 1$: in this case, $h^1_{\text{local}} = A_0$ and $h^1_{\text{hop}} = B_0$, such that (for real-valued $\Delta_0$)
\begin{equation}
\label{eq:HKitaevBdG}
H = \frac{1}{2\pi} \int_{\mathbb{T}} \ud k \, \vec{d}(k) \cdot \vec{\tau}
\end{equation}
with $\vec{d}(k) = \left( 0, \Delta_0 \sin(k), \frac{\mu_0}{2} - t_0 \cos(k) \right)$ and $\vec{\tau}$ being the Nambu pseudospin.

From a slightly different perspective, for the special case $A_0 = \dots = A_{\ell-1}$ and $B_0 = \dots = B_{\ell-1}$ we may again use Fourier transformation applied to the orbitals:
\begin{equation}
\hat{\chi}^{\ell}_{k,\alpha} = \begin{pmatrix} \chatann_{k,\alpha} \\ \chatdag_{-k,\alpha} \end{pmatrix} = \frac{1}{\ell} \sum_{q=0}^{2\pi(\ell-1)} \e^{i \alpha (k + q) / \ell} \begin{pmatrix} \chatann_{(k + q) / \ell} \\ \chatdag_{-(k + q) / \ell} \end{pmatrix}.
\end{equation}
If this is inserted into \eqref{eq:HKitaev_supercell_k} it results in
\begin{equation}
\begin{split}
H &= \frac{1}{2\pi} \int_{\mathbb{T}} \ud k \, \frac{1}{\ell} \sum_{q=0}^{2\pi(\ell-1)} \begin{pmatrix} \chatdag_{(k + q) / \ell} & \chatann_{-(k + q) / \ell} \end{pmatrix} \\
&\qquad \times \left[ A_0 + \big(\e^{i (k + q) / \ell} B_0 + \text{h.c.}\big) \right] \begin{pmatrix} \chatann_{(k + q) / \ell} \\ \chatdag_{-(k + q) / \ell} \end{pmatrix},
\end{split}
\end{equation}
and with the substitution $p = (k + q)/\ell \in \mathbb{T}$, one arrives at
\begin{equation}
H = \frac{1}{2\pi} \int_{\mathbb{T}} \ud p \, \begin{pmatrix} \chatdag_p & \chatann_{-p} \end{pmatrix} \left[ A_0 + \big(\e^{i p} B_0 + \text{h.c.}\big) \right] \begin{pmatrix} \chatann_p \\ \chatdag_{-p} \end{pmatrix}.
\end{equation}
Thus we have again recovered \eqref{eq:HKitaevBdG} (for real-valued $\Delta$), as expected.

\pagebreak


\begin{thebibliography}{10}

\bibitem{BlochDalibardZwerger2008}
I.~Bloch, J.~Dalibard, and W.~Zwerger,
  \href{http://dx.doi.org/10.1103/RevModPhys.80.885}{{Many-body physics with
  ultracold gases}}, {\em Rev. Mod. Phys.} 80, 885--964 (2008).

\bibitem{GoldmanBudichZoller2016}
N.~Goldman, J.~C. Budich, and P.~Zoller,
  \href{http://dx.doi.org/10.1038/nphys3803}{{Topological quantum matter with
  ultracold gases in optical lattices}}, {\em Nat. Phys.} 12, 639--645 (2016).

\bibitem{BlattRoosReview2012}
R.~Blatt and C.~F. Roos, \href{http://dx.doi.org/10.1038/nphys2252}{{Quantum
  simulations with trapped ions}}, {\em Nat. Phys.} 8, 277--284 (2012).

\bibitem{Jurcevic2014}
P.~Jurcevic, B.~P. Lanyon, P.~Hauke, C.~Hempel, P.~Zoller, R.~Blatt, and C.~F.
  Roos, \href{http://dx.doi.org/10.1038/nature13461}{{Quasiparticle engineering
  and entanglement propagation in a quantum many-body system}}, {\em Nature}
  511, 202--205 (2014).

\bibitem{Monroe2017}
J.~Zhang, G.~Pagano, P.~W. Hess, A.~Kyprianidis, P.~Becker, H.~Kaplan, A.~V.
  Gorshkov, Z.~X. Gong, and C.~Monroe,
  \href{http://dx.doi.org/10.1038/nature24654}{{Observation of a many-body
  dynamical phase transition with a 53-qubit quantum simulator}}, {\em Nature}
  551, 601--604 (2017).

\bibitem{Yang2019}
K.~Yang, L.~Zhou, W.~Ma, X.~Kong, P.~Wang, X.~Qin, X.~Rong, Y.~Wang, F.~Shi,
  J.~Gong, and J.~Du,
  \href{http://dx.doi.org/10.1103/PhysRevB.100.085308}{{Floquet dynamical
  quantum phase transitions}}, {\em Phys. Rev. B} 100, 085308 (2019).

\bibitem{YamamotoReview2010}
H.~Deng, H.~Haug, and Y.~Yamamoto,
  \href{http://dx.doi.org/10.1103/RevModPhys.82.1489}{{Exciton-polariton
  Bose-Einstein condensation}}, {\em Rev. Mod. Phys.} 82, 1489--1537 (2010).

\bibitem{Byrnes2014}
T.~Byrnes, N.~Y. Kim, and Y.~Yamamoto,
  \href{http://dx.doi.org/10.1038/nphys3143}{{Exciton-polariton condensates}},
  {\em Nat. Phys.} 10, 803--813 (2014).

\bibitem{PolkovnikovReview2011}
A.~Polkovnikov, K.~Sengupta, A.~Silva, and M.~Vengalattore,
  \href{http://dx.doi.org/10.1103/RevModPhys.83.863}{{Colloquium:
  Nonequilibrium dynamics of closed interacting quantum systems}}, {\em Rev.
  Mod. Phys.} 83, 863--883 (2011).

\bibitem{HeylPolkovnikovKehrein2013}
M.~Heyl, A.~Polkovnikov, and S.~Kehrein,
  \href{http://dx.doi.org/10.1103/PhysRevLett.110.135704}{{Dynamical quantum
  phase transitions in the transverse-field Ising model}}, {\em Phys. Rev.
  Lett.} 110, 135704 (2013).

\bibitem{Karrasch2013}
C.~Karrasch and D.~Schuricht,
  \href{http://dx.doi.org/10.1103/PhysRevB.87.195104}{{Dynamical phase
  transitions after quenches in nonintegrable models}}, {\em Phys. Rev. B}
  87, 195104 (2013).

\bibitem{Canovi2014}
E.~Canovi, P.~Werner, and M.~Eckstein,
  \href{http://dx.doi.org/10.1103/PhysRevLett.113.265702}{{First-order
  dynamical phase transitions}}, {\em Phys. Rev. Lett.} 113, 265702 (2014).

\bibitem{Sharma2015}
S.~Sharma, S.~Suzuki, and A.~Dutta,
  \href{http://dx.doi.org/10.1103/PhysRevB.92.104306}{{Quenches and dynamical
  phase transitions in a nonintegrable quantum Ising model}}, {\em Phys. Rev.
  B} 92, 104306 (2015).

\bibitem{Heyl2015}
M.~Heyl, \href{http://dx.doi.org/10.1103/PhysRevLett.115.140602}{{Scaling and
  universality at dynamical quantum phase transitions}}, {\em Phys. Rev.
  Lett.} 115, 140602 (2015).

\bibitem{VajnaDora2015}
S.~Vajna and B.~D\'ora,
  \href{http://dx.doi.org/10.1103/PhysRevB.91.155127}{{Topological
  classification of dynamical phase transitions}}, {\em Phys. Rev. B}
  91, 155127 (2015).

\bibitem{BudichHeyl2016}
J.~C. Budich and M.~Heyl,
  \href{http://dx.doi.org/10.1103/PhysRevB.93.085416}{{Dynamical topological
  order parameters far from equilibrium}}, {\em Phys. Rev. B} 93, 085416 (2016).

\bibitem{Zvyagin2016}
A.~A. Zvyagin, \href{http://dx.doi.org/10.1063/1.4969869}{{Dynamical quantum
  phase transitions (Review Article)}}, {\em Low Temperature Physics}
  42, 971--994 (2016).

\bibitem{Jurcevic2017}
P.~Jurcevic, H.~Shen, P.~Hauke, C.~Maier, T.~Brydges, C.~Hempel, B.~P. Lanyon,
  M.~Heyl, R.~Blatt, and C.~F. Roos,
  \href{http://dx.doi.org/10.1103/PhysRevLett.119.080501}{{Direct observation
  of dynamical quantum phase transitions in an interacting many-body system}},
  {\em Phys. Rev. Lett.} 119, 080501 (2017).

\bibitem{Flaschner2018}
N.~Fl{\"a}schner, D.~Vogel, M.~Tarnowski, B.~S. Rem, D.~S. L{\"u}hmann,
  M.~Heyl, J.~C. Budich, L.~Mathey, K.~Sengstock, and C.~Weitenberg,
  \href{http://dx.doi.org/10.1038/s41567-017-0013-8}{{Observation of dynamical
  vortices after quenches in a system with topology}}, {\em Nature Physics}
  14, 265--268 (2018).

\bibitem{Heyl2018}
M.~Heyl, \href{http://dx.doi.org/10.1088/1361-6633/aaaf9a}{{Dynamical quantum
  phase transitions: a review}}, {\em Rep. Prog. Phys.} 81, 054001 (2018).

\bibitem{Zunkovic2018}
B.~\v{Z}unkovi\v{c}, M.~Heyl, M.~Knap, and A.~Silva,
  \href{http://dx.doi.org/10.1103/PhysRevLett.120.130601}{{Dynamical quantum
  phase transitions in spin chains with long-range interactions: Merging
  different concepts of nonequilibrium criticality}}, {\em Phys. Rev. Lett.}
  120, 130601 (2018).

\bibitem{Yang2018}
C.~Yang, L.~Li, and S.~Chen,
  \href{http://dx.doi.org/10.1103/PhysRevB.97.060304}{{Dynamical topological
  invariant after a quantum quench}}, {\em Phys. Rev. B} 97, 060304 (2018).

\bibitem{Qiu2018}
X.~Qiu, T.-S. Deng, G.-C. Guo, and W.~Yi,
  \href{http://dx.doi.org/10.1103/PhysRevA.98.021601}{{Dynamical topological
  invariants and reduced rate functions for dynamical quantum phase transitions
  in two dimensions}}, {\em Phys. Rev. A} 98, 021601 (2018).

\bibitem{Sedlmayr2018}
N.~Sedlmayr, P.~Jaeger, M.~Maiti, and J.~Sirker,
  \href{http://dx.doi.org/10.1103/PhysRevB.97.064304}{{Bulk-boundary
  correspondence for dynamical phase transitions in one-dimensional topological
  insulators and superconductors}}, {\em Phys. Rev. B} 97, 064304 (2018).

\bibitem{Fisher1967}
M.~E. Fisher, \href{http://dx.doi.org/10.1088/0034-4885/30/2/306}{{The theory
  of equilibrium critical phenomena}}, {\em Rep. Prog. Phys.} 30, 615--730 (1967).

\bibitem{Pancharatnam1956}
S.~Pancharatnam, \href{http://dx.doi.org/10.1007/BF03046050}{{Generalized
  theory of interference, and its applications}}, {\em Proceedings of the
  Indian Academy of Sciences - Section A} 44, 247--262 (1956).

\bibitem{SamuelBhandari1988}
J.~Samuel and R.~Bhandari,
  \href{http://dx.doi.org/10.1103/PhysRevLett.60.2339}{{General setting for
  Berry's phase}}, {\em Phys. Rev. Lett.} 60, 2339--2342 (1988).

\bibitem{HasanKaneReview}
M.~Z. Hasan and C.~L. Kane,
  \href{http://dx.doi.org/10.1103/RevModPhys.82.3045}{{Colloquium: Topological
  insulators}}, {\em Rev. Mod. Phys.} 82, 3045--3067 (2010).

\bibitem{QiZhangReview}
X.-L. Qi and S.-C. Zhang,
  \href{http://dx.doi.org/10.1103/RevModPhys.83.1057}{{Topological insulators
  and superconductors}}, {\em Rev. Mod. Phys.} 83, 1057--1110 (2011).

\bibitem{Guo2019}
X.-Y. Guo, C.~Yang, Y.~Zeng, Y.~Peng, H.-K. Li, H.~Deng, Y.-R. Jin, S.~Chen,
  D.~Zheng, and H.~Fan,
  \href{http://dx.doi.org/10.1103/PhysRevApplied.11.044080}{{Observation of a
  dynamical quantum phase transition by a superconducting qubit simulation}},
  {\em Phys. Rev. Applied} 11, 044080 (2019).

\bibitem{Wang2019}
K.~Wang, X.~Qiu, L.~Xiao, X.~Zhan, Z.~Bian, W.~Yi, and P.~Xue,
  \href{http://dx.doi.org/10.1103/PhysRevLett.122.020501}{{Simulating dynamic
  quantum phase transitions in photonic quantum walks}}, {\em Phys. Rev.
  Lett.} 122, 020501 (2019).

\bibitem{Xu2018}
X.-Y. Xu, Q.-Q. Wang, M.~Heyl, J.~C. Budich, W.-W. Pan, Z.~Chen, M.~Jan,
  K.~Sun, J.-S. Xu, Y.-J. Han, C.-F. Li, and G.-C. Guo,
  \href{https://arxiv.org/abs/1808.03930}{{Measuring a dynamical topological
  order parameter in quantum walks}}, {\em arXiv e-prints}
  arXiv:1808.03930 (2018).

\bibitem{Tian2018}
T.~Tian, Y.~Ke, L.~Zhang, S.~Lin, Z.~Shi, P.~Huang, C.~Lee, and J.~Du,
  \href{http://dx.doi.org/10.1103/PhysRevB.100.024310}{{Observation of
  dynamical phase transitions in a topological nanomechanical system}}, {\em
  Phys. Rev. B} 100, 024310 (2019).

\bibitem{HuangBalatsky2016}
Z.~Huang and A.~V. Balatsky,
  \href{http://dx.doi.org/10.1103/PhysRevLett.117.086802}{{Dynamical quantum
  phase transitions: Role of topological nodes in wave function overlaps}},
  {\em Phys. Rev. Lett.} 117, 086802 (2016).

\bibitem{Abe1967b}
R.~Abe, \href{http://dx.doi.org/10.1143/PTP.38.72}{{Note on the critical
  behavior of Ising ferromagnets}}, {\em Prog. Theor. Phys.} 38, 72 (1967).

\bibitem{Abe1967c}
R.~Abe, \href{http://dx.doi.org/10.1143/PTP.38.322}{{Singularity of specific
  heat in the second order phase transition}}, {\em Prog. Theor. Phys.}
  38, 322 (1967).

\bibitem{Suzuki1967}
M.~Suzuki, \href{http://dx.doi.org/10.1143/PTP.38.1243}{{Note on the
  singularity of specific heat in the second order phase transition}}, {\em
  Prog. Theor. Phys.} 38(6), 1243 (1967).

\bibitem{Grossmann1969a}
S.~Grossmann and W.~Rosenhauer,
  \href{http://dx.doi.org/10.1007/BF01392423}{{Phase transitions and the
  distribution of temperature zeros of the partition function}}, {\em Z.
  Phys.} 218, 437 (1969).

\bibitem{Grossmann1969b}
S.~Grossmann and V.~Lehmann, \href{http://dx.doi.org/10.1007/BF01392424}{{Phase
  transitions and the distribution of temperature zeros of the partition
  function. II. Applications and examples}}, {\em Z. Phys.} 218, 449 (1969).

\bibitem{Saarloos1984}
W.~van Saarloos and D.~A Kurtze,
  \href{http://dx.doi.org/10.1088/0305-4470/17/6/026}{{Location of zeros in the
  complex temperature plane: absence of Lee-Yang theorem}}, {\em J. Phys. A:
  Math. Gen.} 17, 1301 (1984).

\bibitem{HeylBudich2017}
M.~Heyl and J.~C. Budich,
  \href{http://dx.doi.org/10.1103/PhysRevB.96.180304}{{Dynamical topological
  quantum phase transitions for mixed states}}, {\em Phys. Rev. B} 96, 180304 (2017).

\bibitem{Hofstadter1976}
D.~R. Hofstadter, \href{http://dx.doi.org/10.1103/PhysRevB.14.2239}{{Energy
  levels and wave functions of Bloch electrons in rational and irrational
  magnetic fields}}, {\em Phys. Rev. B} 14, 2239--2249 (1976).

\bibitem{Aidelsburger2013}
M.~Aidelsburger, M.~Atala, M.~Lohse, J.~T. Barreiro, B.~Paredes, and I.~Bloch,
  \href{http://dx.doi.org/10.1103/PhysRevLett.111.185301}{{Realization of the
  Hofstadter Hamiltonian with ultracold atoms in optical lattices}}, {\em Phys.
  Rev. Lett.} 111, 185301 (2013).

\bibitem{Kitaev2001}
A.~Y. Kitaev, \href{http://dx.doi.org/10.1070/1063-7869/44/10S/S29}{{Unpaired
  Majorana fermions in quantum wires}}, {\em Phys. Usp.} 44, 131 (2001).

\bibitem{Beenakker2013}
C.W.J. Beenakker,
  \href{http://dx.doi.org/10.1146/annurev-conmatphys-030212-184337}{{Search for
  Majorana fermions in superconductors}}, {\em Annu. Rev. Condens. Matter
  Phys.} 4, 113--136 (2013).

\bibitem{Yang2017}
C.~Yang, Y.~Wang, P.~Wang, X.~Gao, and S.~Chen,
  \href{http://dx.doi.org/10.1103/PhysRevB.95.184201}{{Dynamical signature of
  localization-delocalization transition in a one-dimensional incommensurate
  lattice}}, {\em Phys. Rev. B} 95, 184201 (2017).

\bibitem{Yin2018}
H.~Yin, S.~Chen, X.~Gao, and P.~Wang,
  \href{http://dx.doi.org/10.1103/PhysRevA.97.033624}{{Zeros of Loschmidt echo
  in the presence of Anderson localization}}, {\em Phys. Rev. A} 97, 033624 (2018).

\bibitem{Srivastav2018}
V.~Srivastav, U.~Bhattacharya, and A.~Dutta,
  \href{http://dx.doi.org/10.1103/PhysRevB.100.144203}{{Dynamical quantum phase
  transitions in extended toric-code models}}, {\em Phys. Rev. B} 100, 144203
  (2019).

\end{thebibliography}
\end{document}